# The rapid formation of a large rotating disk galaxy three billion years after the Big Bang[1]

*Nature in press (August 06)*

R.Genzel[1,2], L.J.Tacconi[1], F.Eisenhauer[1], N.M. Förster Schreiber[1], A.Cimatti[3,1],

E.Daddi[4], N.Bouché[1], R.Davies[1], M.D.Lehnert[1], D.Lutz[1], N.Nesvadba[1], A.Verma[1],

R.Abuter[1], K.Shapiro[5], A.Sternberg[6], A.Renzini[7], X.Kong[8], N.Arimoto[9] & M.Mignoli[10]

[1] *Max-Planck-Institut für extraterrestrische Physik (MPE), Giessenbachstr.1, 85748 Garching, Germany ( genzel@mpe.mpg.de)*

[2] *Department of Physics, Le Conte Hall, University of California, 94720 Berkeley, USA*

[3] *Istituto Nazionale di Astrofisica – Osservatorio Astrofisico di Arcetri, Largo E. Fermi 5, I-50125 Firenze, Italy*

[4] *Spitzer Fellow, National Optical Astronomy Observatory, 950 North Cherry Avenue, Tucson, AZ 85719, USA*

[5] *Department of Astronomy, Campbell Hall, University of California, Berkeley, 94720, USA*

[6] *School of Physics and Astronomy, Tel Aviv University, Tel Aviv 69978, Israel*

[7] *Osservatorio Astronomico di Padova, Vicolo dell'Osservatorio 5, Padova, I-35122, Italy*

[8] *Center for Astrophysics, University of Science and Technology of China, Hefei, 230026, PR China*

[9] *National Astronomical Observatory, Mitaka, Tokyo 181 8588, Japan*

[10] *Istituto Nazionale di Astrofisica – Osservatorio Astronomico di Bologna, Via Gobetti 101, I-40129 Bologna, Italy*

**Over the past two decades observations and theoretical simulations have established a global frame-work of galaxy formation and evolution in the young Universe (1-3). Galaxies formed as baryonic gas cooled at the centres of collapsing dark matter halos. Mergers of halos and galaxies led to the hierarchical build-up of galaxy mass. It remains unclear, however, over what timescales galaxies were assembled and when and how bulges and disks, the primary components of present day galaxies, were formed. It is also puzzling that the most massive galaxies were more abundant and were forming stars more rapidly at early epochs than expected from models (4-7). A major step forward in understanding these issues requires well resolved physical information on individual galaxies at high redshift. Here we report adaptive optics, spectroscopic observations of a representative luminous star forming galaxy when the Universe was only twenty percent of its current age. The far superior angular resolution of these data compared to our first study (8) reveals the physical and dynamical properties of a high redshift galaxy in unprecedented detail. A large and massive rotating proto-disk is channelling gas toward a growing central stellar bulge hosting an accreting massive black hole. The high gas surface densities, large star formation rate and moderately young stellar ages suggest rapid assembly, fragmentation and conversion to stars of an initially very gas rich proto-disk, with no obvious evidence for a major merger.**

Imaging spectroscopy of high redshift galaxies at high angular resolution of well understood rest-frame optical spectral diagnostics is now becoming feasible with advanced instruments on large ground-based telescopes. This promises new empirical information about the crucial epoch of galaxy evolution near cosmological redshift $z$~2, about 3 billion years after the Big Bang when the Universe was about 20% of its current age. We have recently begun a study of a representative sample of $z$~2-3 star forming galaxies, selected based on their rest-frame ultra-violet/optical fluxes and colours, with the near-infrared integral field spectrometer SINFONI on the Very Large Telescope of the European Southern Observatory (**9, 10**). Our first results (**8**) revealed that fairly large and massive proto-disk galaxies were present already at $z$~2-3. We did not have sufficient resolution, however, to distinguish unambiguously between a merger and disk interpretation, or to resolve the bulge and disk components. For one of these luminous star forming galaxies, BzK-15504 ($z$=2.38: **11, 12**), the presence of a nearby star and excellent atmospheric conditions now allowed us, for the first time, to take full advantage of the adaptive optics mode of SINFONI. We achieved an angular resolution of ~0.15" (1.2 kpc or 4000 light years), more than three times better than in our previous work. BzK-15504 is a fairly typical representative of rest-frame optically bright, actively star forming galaxies at that redshift (for details see caption of Fig.1 and Supplementary Information). The SINFONI spectral data reveal the spatial distribution and kinematics of spectroscopic tracers in unprecedented detail. Hα line emission tracing ionized gas in many compact star formation complexes is distributed over an extended region of diameter ~1.6" (13 kpc), approximately centred on the continuum peak (Fig.1). The Hα surface brightness distribution is somewhat asymmetric (the emission toward the north-west is significantly brighter) but low level Hα emission is detected over a comparable area in both north-west and south-east quadrants. Similarly asymmetric distributions of the bright ionized gas are often seen in otherwise symmetric, local Universe disk galaxies. There they reflect the instantaneous distribution of the youngest star forming regions and, in some cases, of dust extinction (**13, 14**). The velocity field of Hα is remarkably symmetric and is described very well by a combination of circular motion (rotation) in the outer parts and strong radial flow in the nuclear region. BzK-15504 also has a bright near-nuclear concentration of Hα emission, accompanied by wider lines of 450 km/s FWHM (Figs.1/2, Supplementary Information).

In the outer regions ($r$>0.4") the highest and lowest velocity HII region complexes are found at position angle ~24° west of north, which is also the morphological major axis. Along this axis, the absolute values of the mean line velocities (relative to the systemic velocity) increase sharply with distance from the centre and reach a constant value at $|r|$>0.4" (Fig.2b). Along the minor axis in the blue-shifted north-western quadrant the absolute values of the mean velocities decrease smoothly on either side of the major axis (Fig.1 b-i, Fig.2d). Taken together these characteristics constitute compelling qualitative evidence for rotation in a disk inclined with respect to the sky plane (**15**). For quantitative modelling we adopted an azimuthally symmetric, exponential disk (see Supplementary Information), which fits the data very well. The best fitting models from $\chi^2$-minimisation (Table 1) have a radial exponential scale length of 4.5 kpc, a maximum circular velocity of 230 km/s and an overall dynamical mass of $1.1 \times 10^{11}$ M$_\odot$ within $r$~8 kpc. While the ionized gas in BzK-15504 thus is clearly a large rotating disk, the question arises whether this conclusion also holds for the underlying galaxy as a whole, or whether a merger configuration might not also be possible. The possibility of a major merger (mass ratio of the two galaxies <3:1) is not compatible with the observed kinematics. In the case of two galaxy mass centres separated on the same scale as the gas disk ($\geq$ 4 kpc), the gas dynamics would be strongly perturbed and substantially deviate from that of simple disk. In the case of a very advanced major merger, most of the gas would have small angular momentum and would be concentrated on the spatial scale of the (stellar) merger remnant (<1 kpc) even if a small fraction of loosely bound gas may be in a rotating disk with large angular momentum. Finally, in a minor merger (mass ratio >3:1) the more massive galaxy, including its gas distribution, can retain a significant angular momentum and resemble a rotating disk. This case is compatible with our observations. We conclude that BzK-15504 is a large and massive, proto-disk galaxy with an angular momentum comparable to local Universe Sb/c spirals.

Several properties of the source clearly indicate that the proto-disk was assembled rapidly, on a time scale of a few hundred million years, and has been converting a significant fraction of its total baryonic mass to stars on that time scale, with no obvious evidence for a major merger. The dynamical properties and the global and local matter surface densities (Table 1) show that the gas in the proto-disk is unstable to global star formation ('Toomre' $Q$ parameter $\leq$1, **16, 17**) and fragmentation (**18**). If the local gas motions seen toward the star forming complexes in Fig.1 are virialised (Fig.3 caption, Supplementary Information) the characteristic diameters (~0.1-0.2") and intrinsic local velocity dispersions of ~30-60 km s$^{-1}$ imply very large complex masses of 2-10 10$^8$ M$_{\odot}$, which in turn are exactly what would be expected for a globally unstable disk. From the Hα line flux and a 'Kennicutt' (**19**) conversion from Hα luminosity to star formation rate and modelling of the UV spectral energy distribution (Supplementary Information) we infer a total star formation rate of 100-200 $M_\odot$ yr$^{-1}$ for a Chabrier/Kroupa (**20,21**) stellar mass function (Table 1). This is close to the 'maximum' rate (**22**) for forming the current stellar mass of ~8 $10^{10}$ $M_\odot$ (Table 1) on a time scale of ~500 Myrs, the stellar age derived from a fit to the spectral energy distribution of the object (see Supplementary Information). Further, the ratio of circular velocity to local gas velocity dispersion in BzK-15504 is $v_c/\sigma$~2-4 (all corrected for instrumental effects and inclination), as in three other such galaxies in (**8**) (see Supplementary Information). The thin disks of local spiral galaxies have much greater ratios: $v_c/\sigma$~10-50. The ionized gas disk is dynamically 'hot'. The energy source for this large dispersion is plausibly due to 'feedback' through supernova explosions, or radiation and winds from the active star formation itself, or due to the tapping of the accretion energy in the disk's assembly (**8, 23-25**). All scenarios require a fairly rapid (a few hundred Myrs) formation time scale, consistent with the arguments given above. As discussed above the remarkable feature of BzK-15504 is that in this case the symmetric properties of the observed outer velocity field argue against a recent major merger event of two galaxies as the trigger of this intense star formation activity. Our data imply a more quiescent but still fast gas inflow from the dark halo, or a rapid series of minor mergers. Combining the gas and stellar masses from Table 1 fully accounts for the inferred dynamical mass within ~10 kpc. The disk appears to be dominated by baryons.

While the outer disk is well described by circular motion, velocity field and data-model difference maps of the inner disk show strong deviations from rotation within ~0.4" (3 kpc) of the nucleus. This conclusion is apparent when inspecting the residual maps in Fig. 3, which exhibit significant velocity residuals south-west of the nucleus. The origin of these residuals is gas at slightly blue-shifted velocities entering the nuclear region from the north-west, then passing along the minor axis below the nuclear position and exiting south-east into the red-shifted part of the disk (see Fig.1a, d-g, visualised by dotted S-shaped curve). This trend is reflected in the overall velocity field of Fig.3a as a twist of the iso-velocity contours, reaching an amplitude of 70 to 120 km s$^{-1}$ in the velocity residual map (Fig.3c). Very similar patterns are seen in many local Universe galaxies (e.g. **26, 15**), where they are interpreted as the tell-tale signature of radial inward streaming gas from the disk into the nucleus (e.g. in response to a stellar bar). This inflow could be an important contributor to the growth of a substantial bulge whose presence is already apparent as a bright stellar peak near the centre of BzK-15504 (see Supplementary Information). In addition there is a red-shifted high velocity tail (to ~500 km s$^{-1}$) of the Hα emission ~0.2"east of the nuclear continuum peak (Fig.1j). This feature is also prominent in [NII] (Supplementary Information) and may be powered by hard ultra-violet radiation from an active galactic nucleus (AGN), whose presence in BzK-15504 at that position is indicated by the properties of its UV spectrum (Supplementary Information). Inward nuclear flow and AGN contribution plausibly explain the relatively large nuclear velocity dispersion (Figure 2c) that is not accounted for by the disk rotation model.

Our observations suggest a self-consistent picture of rapid gas inflow from the halo, followed by gravitational instability and star formation in a massive, gas rich proto-disk, and subsequent build-up of a central bulge through inflow triggered by disk instabilities and/or minor mergers (**18, 27**). While BzK-15504 is the only z~2 galaxy for which such high resolution data are presently available, its global properties, including star formation rate, stellar and dynamical masses and size, are similar to the general population of actively star-forming galaxies at z~2 (Supplementary Information). As such, the phenomena observed in this galaxy, such as rapid gas inflow, gravitational instability, and subsequent build-up of a central bulge are likely of general importance for the rapid formation and assembly of massive ($\sim M_*$) galaxies in this critical cosmic epoch (**6,7**). If the large thickness of the Hα gas layer inferred from the observed $v_c/\sigma$~3 ratio is characteristic of the underlying stellar disk, an obvious question raised by our study is whether we are witnessing the formation of a 'thick' disk, possibly related to the so called 'old thick' stellar disks seen in a number of spiral galaxies at low redshift (**28**). Alternatively the z~2 proto-disks may be later destroyed by major mergers and contribute to forming elliptical/S0 galaxies.

Acknowledgments. We would like to thank the ESO and MPE SINFONI team members for their high quality work, which made these technically difficult observations possible. We also thank the Paranal staff, especially Julio Navarrete and Paola Amico for their excellent support. We are grateful to Andreas Burkert, Ortwin Gerhard and Pierluigi Monaco for interesting discussions. A.C. gratefully acknowledges support through a 'Bessel Prize' of the Alexander von Humboldt Foundation.

The authors have no competing financial interests.

Correspondence and requests for materials should be addressed to R.Genzel (genzel@mpe.mpg.de).

## Table 1.  Physical Properties of BzK-15504[2]

| | | |
|---|---|---|
| $z$ | 2.3834 | look-back time 10.7 Gyrs, 1"≡8.135 kpc |
| $K_s$ | 19.2 | total $K_s$ band magnitude (Vega system, uncorrected for 30% line emission) |
| $A_V$ | 0.9 (+0.3,-0.3) | extinction derived from spectral energy distribution fitting (SED) |
| $R_*$ | 140 (+110,-80) $M_\odot$ yr$^{-1}$ | star formation rate from Hα flux of $2.5 \times 10^{-16}$ erg s$^{-1}$ cm$^{-2}$ (**19**) and UV SED, for IMF in (**20,21**) and $A_V$=0.9 mag |
| $t_*$ | 5 (+5,-2) $10^8$ yr | stellar age from SED fitting, for continuous or 3 $10^8$ yr duration burst |
| $\Sigma_*$ | 1.2 $M_\odot$ yr$^{-1}$ kpc$^{-2}$ | star formation rate surface density |
| $M_{dyn}$ | 11.3±1 $10^{10}$ $M_\odot$ | dynamical mass within r=1.1", corrected for inclination $i$=48±3$^0$ |
| $M_*$ | 7.7  (+3.9,-1.3) $10^{10}$ $M_\odot$ | from SED and IMF from (**20,21**), excluding stellar mass loss |
| $M_{gas}$ | 4.3 $10^{10}$ $M_\odot$ | total gas mass from Hα and Schmidt-Kennicutt law (**19**), for $A_V$=0.9 |
| $\Sigma_{gas}$ | 350 $M_\odot$ pc$^{-2}$ | total gas surface density from Hα surface brightness and Schmidt-Kennicutt law (**19**) |
| $v_c$ | 230±16 km s$^{-1}$ | circular velocity at $r$=5-10 kpc |
| $R_{1/e}$ | 4.5±1 kpc | radial scale length of Hα disk |
| $z_{1/e}$ | 1±0.5 kpc | vertical scale length of Hα disk, from $v_c/\sigma$=3±1 |
| $Q$ | 0.8±0.4 | Toomre Q-parameter for global disk stability (**16**) |

---

[2] See Supplementary Information for details

Figure 1. Velocity maps of Hα emission in BzK-15504 ($z$=2.3834). Panel a is an RGB composite colour map of the entire Hα emission, with three colours encoding the total velocity range of the emission. Panels (b-i) are velocity maps summed over 65 km/s. Panel j (260 km/s range) displays the highly red-shifted feature east of the nucleus. The velocity maps b-e show that the gas with the largest blueshift is found near the major axis but as the velocity becomes less blue the emission moves further and further away on either side of this axis. This pattern is a clear characteristic of a rotating disk. The crosses in the panels mark the position of the continuum peak. The dotted, thin white curve outlines the shape of the integrated Hα emission at the ~20% level. The dotted, light blue, S-shape visualises the radial flow discussed in the text. The two yellow lines mark the orientation of the major axis of the disk. North is up and East to the left, and a bar marks the length of 0.5".The source BzK-15504 was discovered as part of a K-band (1.9-2.4μm, rest-frame ~6500 Å) survey (to $K_{AB}$<21.9) in the $R.A.$=11[h], $Dec.$=-21[0] 'deep 3a-F' Subaru field (**12,** Supplementary Information). Star forming galaxies at $z$~2 were selected from the so-called '(s)BzK' colour criteria (Supplementary Information). Follow-up spectroscopy was then carried out with VIMOS on the ESO VLT. The source BzK-15504 ($K_{AB}$=21.08, $z$=2.38) is among the brightest 30% of the $z$≥2 BzK galaxies. The rest frame UV spectrum of the source (Fig2a of Supplementary Information) exhibits strong Lyα, CIV 1549 Å and CIII] 1909 Å emission, indicative of the presence of a central AGN. The SINFONI data were taken in two sets of six-hour integrations in the K-band, both using a nearby $V$=16.3 star as reference for the MACAO adaptive optics system. SINFONI delivers spectra simultaneously over a contiguous two dimensional field with 64x32 pixels. The FWHM spectral resolution is 54 km/s. The data set shown in Figures 1-3 was taken in excellent atmospheric seeing (FWHM on the visible guider ~0.4-0.6") and coherence time (7-10 ms) conditions and used the 0.05"x0.1" pixel scale. The resulting FWHM angular resolution is 0.15", corresponding to1.2 kpc (4000 light years). Another data set was taken in the 0.125"x0.25" pixel scale and resulted in a FWHM spatial resolution of 0.45". The SINFONI data were reduced and calibrated with the custom-developed 'SPRED' (**29, 9**) software, with additional tools to treat very faint signals and analyse the final data cubes. All physical units in this paper are based on a concordance, flat $\Lambda CDM$ cosmology with $H_0$=70 km s$^{-1}$ Mpc$^{-1}$.

Figure 2. Velocity and intensity distributions along the major and minor axes of the source. a: Hα intensity along the morphological and kinematic major axis at position angle 24$^°$ west of north. b and c: peak velocity and velocity dispersion and their 1σ uncertainties extracted along the major axis by fitting Gaussians to spectra in selected apertures (marked by the range in position offset in the panels). d: peak velocity along the minor axis (p.a. 114$^°$ west of north) through a position 0.69" north-west of the centre and averaged over 0.25" along the major axis. Filled circles denote spectra in the 0.15" high resolution data of Figure 1, while crosses represent the second independent data set at 0.45" spatial resolution. Continuous curves denote the best fitting exponential disk model to the data. The dotted curve in panel b is the inclination and resolution corrected, intrinsic rotation curve inferred from our modelling.

Figure 3. Two dimensional distributions of first and second moments of the Hα velocity distribution. a: extracted mean velocity map. b: extracted velocity dispersion map. c and d: difference maps of a and b and the corresponding best fitting. exponential disk model distributions  Superposed are contours of integrated Hα emission. In all cases the data and models were smoothed to 0.19" FWHM. The crosses denote the position of the continuum peak. The strong deviations (in the top and bottom, left panels) near the dynamical centre of the velocity field from that of the simple rotation pattern in the outer disk indicate a 70-120 km s$^{-1}$ component of radial motion, either inflow or outflow. The spatial connection of this radially streaming gas to the outer disk apparent from the channel maps in Fig.1 strongly favours radial inflow.

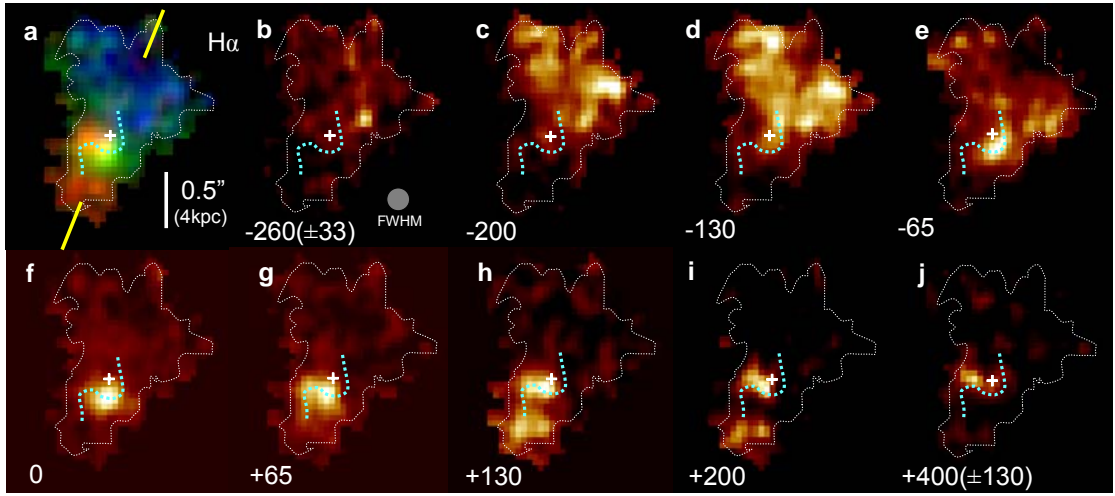

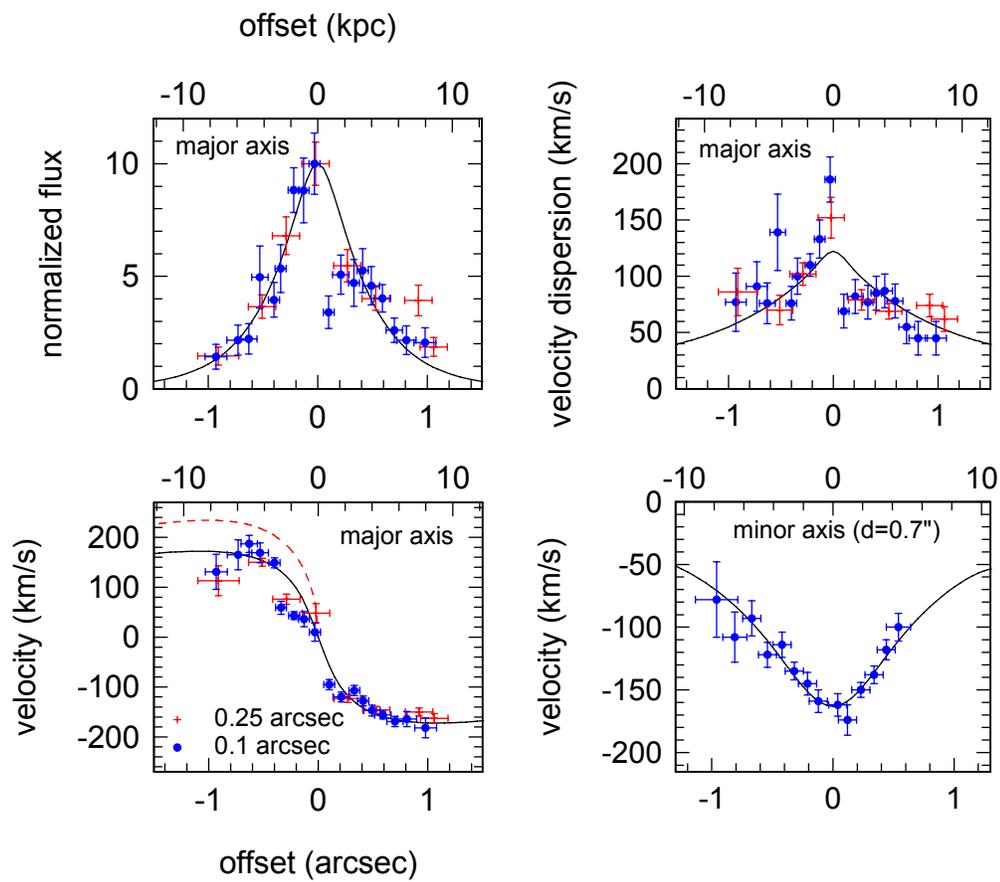

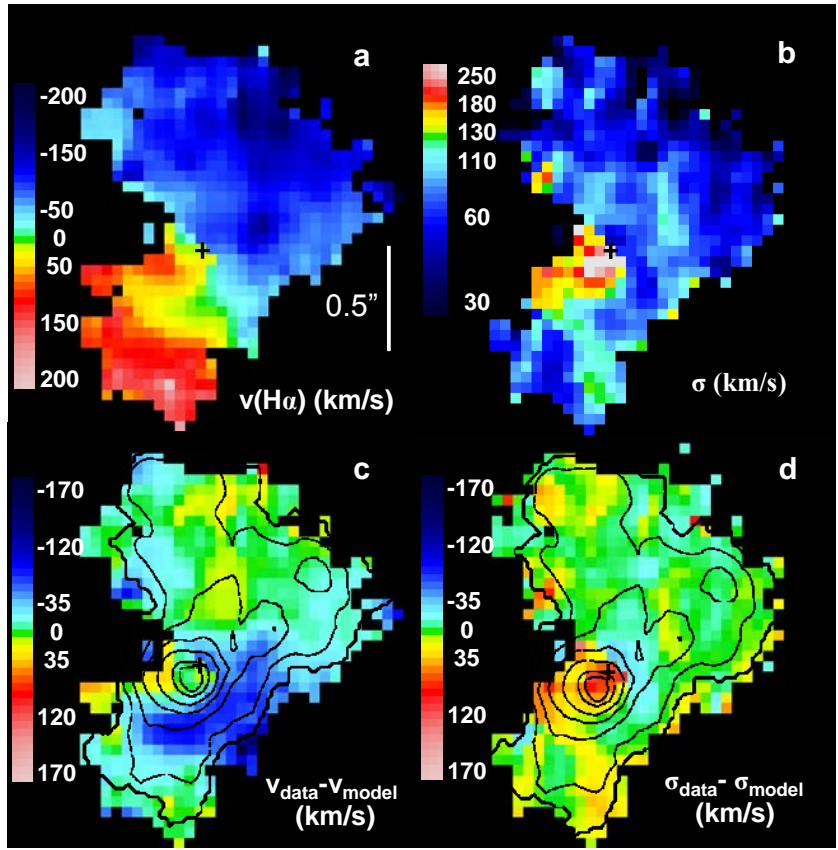

# Supplementary Information

## Source selection

The galaxy presented in this paper was drawn from a K-band (1.9-2.4μm) selected sample with a magnitude limit of $K_s < 20$ (Vega photometric system) in the wide Deep3a field and applying the so-called BzK colour selection (**1, 2**). This method is based on the $z$-$K_s$ versus $B$-$z$ colour-colour diagram, where the broad-band filters $B, z$ and $K_s$ are cantered at the observed wavelengths of about 0.4, 0.9 and 2.2 μm, respectively. This criterion is very efficient at selecting galaxies in the redshift range of $1.4 < z < 2.5$, with the $B$ and $z$ filters sampling the rest-frame spectral region shortward of the Balmer and 4000 Å continuum breaks. With $BzK \equiv (z$-$K_s) - (B$-$z)$ (AB photometric scale), star-forming galaxies can be extracted with $BzK > -0.2$. Compared to optical selection techniques (e.g., the "BM/BX" criteria targeting $1.5 < z < 2.5$ galaxies based on their rest-frame UV properties; **3, 4**), the $BzK$ method has the advantage of being insensitive to dust reddening.

In terms of the overall properties of $BzK$ selected galaxies in the survey of (**2**) BzK-15504 is among the ~30% brightest in terms of K-band magnitude and has a very blue $B$-$z$ colour but is otherwise close to the median values of the full sample in terms of its stellar mass and star formation rate (Fig.S1). In contrast, optically-selected star-forming z~2 galaxies, in particular BM/BX objects, typically are fainter in the K-band, have lower stellar masses (median ~$2 \times 10^{10}$ M$_\odot$), and lower star formation rates (median ~35 M$_\odot$yr$^{-1}$), although the overlap in range of properties with star-forming BzK objects increases for $K_s < 20$ BM/BX galaxies (e.g., **5, 6, 7**). All masses and star formation rates quoted from the literature have been corrected here for the Chabrier/Kroupa IMF (**8, 9**) adopted throughout this paper. Less information exists on the sizes, but BzK-15504 is comparably large to the 9 $K_s < 20$ star-forming BzK objects discussed by (**10**) and many of the z~2 BX galaxies in our first SINFONI study (**11**).

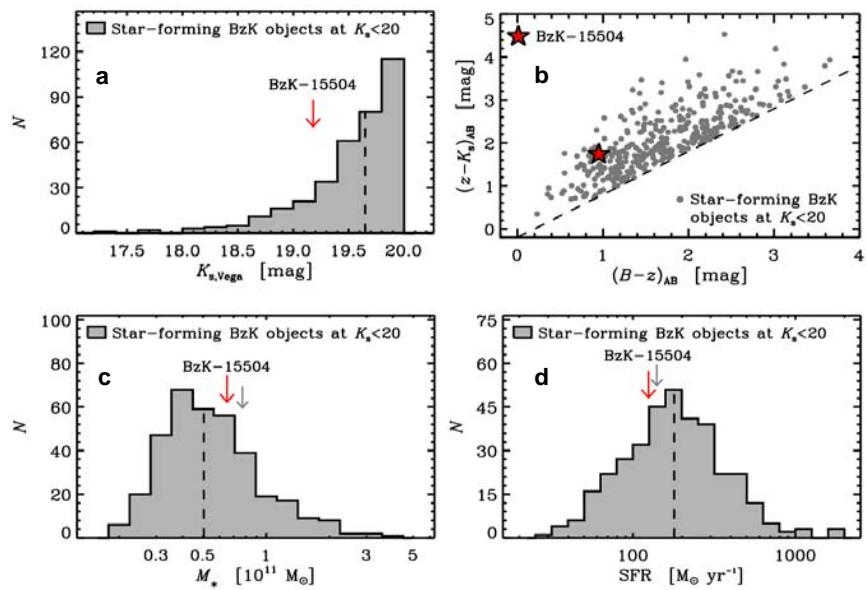

Fig. S1: Histograms of the K-band magnitude (a), colour (b), stellar mass (c) and star formation rate (d) for the star-forming BzK sample in the Deep 3a field to $K_s$=20 mag, from which BzK-15504 was drawn. The stellar mass and star formation rate estimates for this sample were derived from the empirical relationships appropriate for star-forming BzK objects of (**1**) (and corrected for the Chabrier/Kroupa IMF adopted in this paper; **8, 9**). The arrows in each panel indicate the estimates for BzK-15504: the large red arrow corresponds to the value derived using the same empirical relationships as for the full sample for consistent comparison, while the smaller grey arrow indicates the results of the SED modelling as reported in Table 1 (the two estimates of stellar mass and star formation rate differ by about 20%, well within the typical uncertainties of both methods).

Spectra

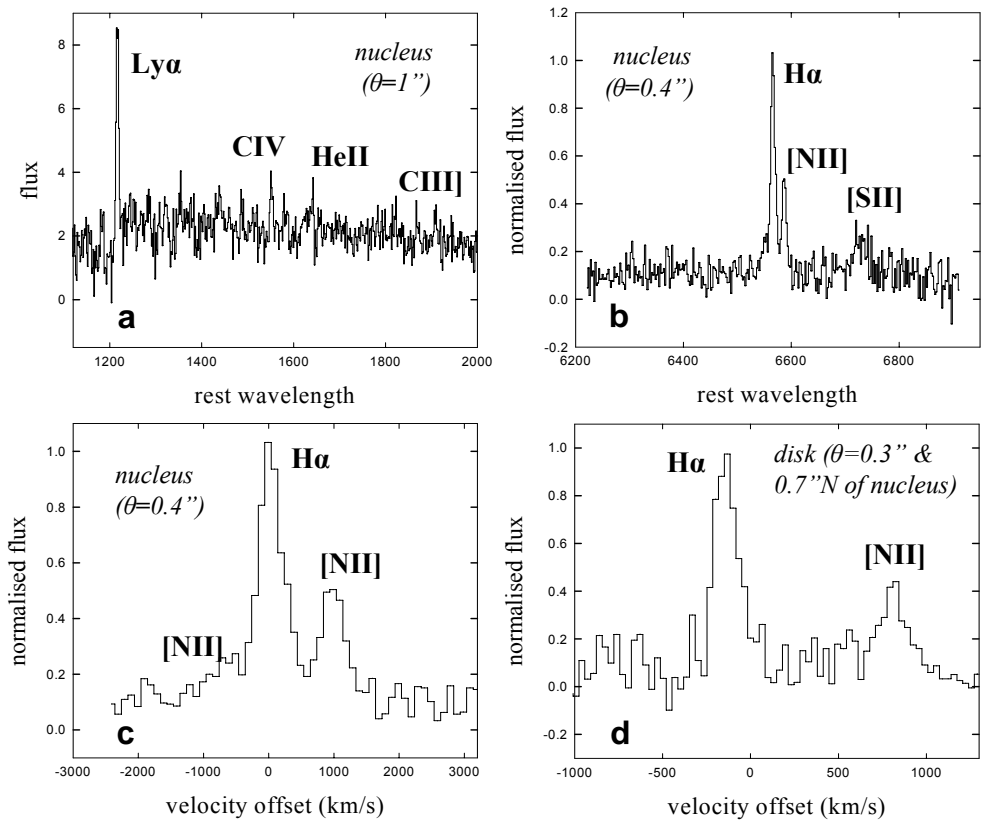

Figure S2. Ultra-violet (UV: Daddi et al. in preparation) and optical, rest-frame spectra of BzK-15504. a: restframe UV spectrum of the nucleus, taken in a 1" slit with VIMOS on the ESO VLT. Emission due to λ=1216 Å HI Lyα, λ=1549 Å CIV, λ=1640 Å HeII and λ=1909 Å CIII] all result in a redshift of 2.38. b: nuclear spectrum in rest-frame R-band extracted from our SINFONI data in a 0.4" diameter aperture. In addition to λ=6565 Å HI Hα, the λ=6583, 6550 Å lines of [NII] and the λ=6718, 6733 Å lines of [SII] are apparent. c: Zoom in on the nuclear spectrum with the velocity scale set at z=2.3834 on Hα. The line profile is fit by a FWHM line width of 450 km s$^{-1}$ and a 6583 [NII]/Hα flux ratio of 0.46. d: Hα/[NII] spectrum of a typical HII complex in the outer disk, extracted in a 0.3" diameter aperture 0.69" north of the nucleus. The spectrum is fit with an intrinsic line width of 145 km s$^{-1}$ and a 6583 [NII]/Hα flux ratio of 0.36.

## K-band continuum/Hα and [NII]/Hα ratio maps

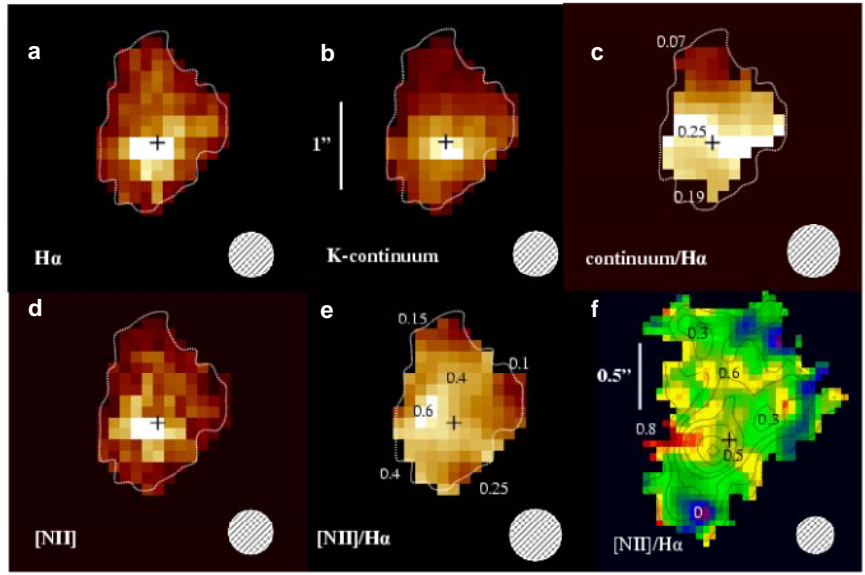

Figure S3. λ=6583 [NII] and line free K-band continuum maps of BzK-15504, compared to Hα. Panels a-e are from the 0.125"x0.25" pixel data set (FWHM resolution 0.45", indicated by shaded circles at the bottom right of each map). Panel f is from the high resolution data set (smoothed to 0.23" FWHM), with contours of integrated Hα superposed. In all maps the region with significant Hα

emission has been masked and the cross denotes the position of the continuum peak. a: integrated Hα emission; b: line free 2.05-2.3μm continuum; c: continuum to Hα flux ratio, with typical values marked. d: integrated [NII] flux map; e/f: [NII]/Hα flux ratio maps, with typical values marked.

## Spectral energy distribution modelling and properties of stellar population

The optical photometry was obtained with the Suprime Camera at the SUBARU telescope in the *B, R, I,* and *z′* bands, and the near-infrared *J* and *Ks* band photometry was collected with SOFI at the ESO New Technology Telescope (**7**). The data are complemented with optical *U* and *V* band photometry obtained as part of the ESO Imaging Survey (EIS) with the WFI instrument at the MPG/ESO 2.2 metre telescope. For the SED modelling, we applied the evolutionary synthesis models of Bruzual & Charlot (**1**), after accounting for the contribution of emission lines to the broad-band fluxes (30% from Hα+[NII] lines in the $K_s$ band, 5% from Lyα in the B band) and for Galactic extinction in the direction of BzK-15504 (E(B-V) = 0.043 mag).  The broad-band fluxes correspond to the total emission of the galaxy, as determined from the near-infrared $K_s$ band emission (Daddi et al., in preparation). For the modelling, we fixed the redshift of BzK-15504 at $z = 2.3834$ and followed the same procedure as in (**5**).  We explored a wide range of ages (from $10^6$ to $3\ 10^9$ yr) and extinction values, and determined the best-fit model from $\chi^2$-minimisation to the broad-band SED. All other model parameters are fixed, including the star formation history, the stellar initial mass function (IMF), the metallicity, and the dust extinction law. For BzK-15504, we adopted solar metallicity, a Calzetti law for dust reddening, a Salpeter IMF between 0.1 and 100 $M_\odot$ and constant star formation rate. We corrected all derived masses and star formation rates for a more realistic Kroupa/Chabrier IMF by dividing them by 1.6 (**2,8**). The uncertainties on the best-fit properties were derived from Monte-Carlo simulations, perturbing the input photometry assuming Gaussian uncertainties of the fluxes. For these parameters, we obtain a best-fit age of 7.2 (+3,-3.6) $10^8$ yr, extinction $A_V$ of 0.9 (+0.3/-0.0) mag, stellar mass of $M_* = 8.2$ (+3.4/-1.8) x $10^{10}$ $M_\odot$, total gas mass consumed in forming stars of $M_{tot}$ = 1.0 (+0.5,-0.2) x $10^{11}$ $M_\odot$, and a current star formation rate $SFR$ = 140 (+110,-10) $M_\odot$ yr$^{-1}$.

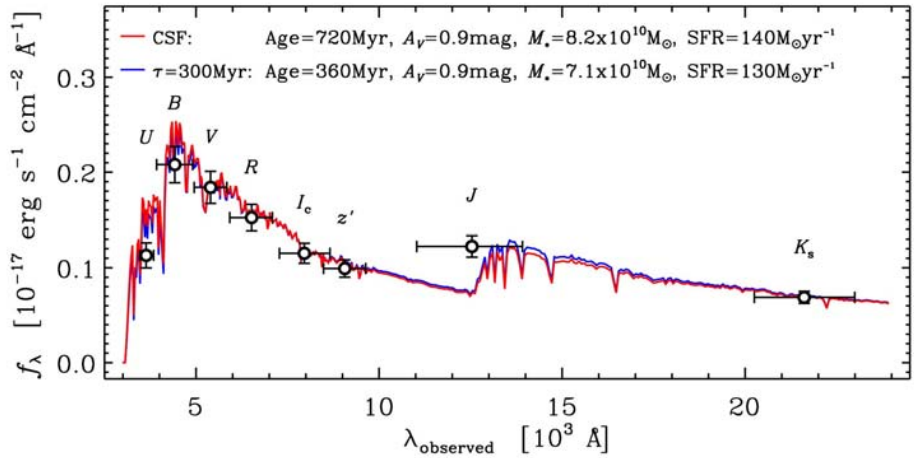

Figure S4. Broad-band spectral energy distribution (SED) of BzK-15504 (large dots), overplotted with the best-fitting spectral synthesis model for two different star formation histories: constant star formation (CSF; red line) and exponentially declining star formation rate with e-folding timescale of $\tau$ = 3 $10^8$ yr (blue line). The best-fit age, visual extinction, stellar mass, and current star formation rate

are indicated for each model in the plot. The vertical error bars indicate the $1\sigma$ uncertainties of the flux measurements, and the horizontal error bars show the width of the different photometric passbands.

We also explored models with an exponentially declining star formation and considered models with sub-solar metallicity and an extinction law appropriate for the Small Magellanic Cloud. While the best-fit ages, extinction, and star formation rates vary significantly for different decay time parameters (ages from 3.6 to 7.2 $10^8$ yr, $A_V$ between 0.3 and 0.9 mag, and star formation rates from 20 to 140 $M_\odot$ yr$^{-1}$ all increasing with decay timescale), the stellar and total masses are better constrained and range from 6-9 x $10^{10}$ $M_\odot$ and 7-11 x $10^{10}$ $M_\odot$. Models with lower metallicity lead to somewhat higher ages, lower extinction values, and lower star formation rates but similar stellar and total masses. The best-fit current star formation rate obtained from the SED modelling is consistent with independent estimates that lie in the range $100 - 200$ $M_\odot$ yr$^{-1}$, based on the H$\alpha$ flux corrected for an extinction of about 1 mag and using the Kennicutt (**6**) conversion between H$\alpha$ luminosity and star formation rate, based on the extinction-corrected rest-frame UV continuum luminosity and the conversion proposed in (**9**), and based on the 24$\mu$m flux (79 $\mu$Jy) using models in (**3**) or the average bolometric SED for $z$~2 star-forming *BzK* objects (**4**).

## Kinematic Modelling

We considered simple models consisting of azimuthally symmetric rotating disks, in which the input light and mass distributions can be arbitrary functions of radius. Other input parameters include the total mass to an outer cutoff radius, the inclination angle - $i$ - of the normal to the disk's plane with respect to the line of sight, the radial scale length and the vertical (z-axis) FWHM thickness of the disk.  From this z-thickness the z-component of the velocity dispersion is computed assuming hydrostatic equilibrium in the limit of a thin and extended disk. The smearing due to the instrumental resolution in the spatial and spectral domains is taken into account by convolving the inclined model with the Gaussian point spread function of the appropriate width. The outputs of each model are full data cubes and velocity moment maps that can be directly compared to the observations (see Figure 3 of the main paper). One-dimensional cuts along any preferred direction can also be obtained and are used to quantitatively constrain the fit parameters (Figure 2).

For determining the disk parameters and their uncertainties in Table 1 we adopted an exponential mass surface distribution of outer radius 1.6". We determined the position angle of the major axis (24° west of north) by searching for the largest amplitude in mean velocity along slits passing through the nominal continuum centre position. We then varied inclination, total mass, radial and z-scale lengths to reach minimum $\chi^2$ with respect to the observed one dimensional distributions in total Hα intensity, mean velocity and velocity dispersion along the major axis and also in Hα mean velocity along the minor axis (114° west of north) in the undisturbed and well sampled north-western section of the disk, ~0.69" from the nucleus. Input-errors in the $\chi^2$ evaluation were the 1σ fit uncertainties of the line profiles. The Hα major axis intensity distribution is most sensitive to the radial scale length (best fit $R_{1/e}$= 0.56"±0.12", all uncertainties given here are 68% confidence). The mean velocity distribution along the major axis is most sensitive to total mass for a given inclination and the mean velocity distribution along the minor axis is most sensitive to inclination. The combined $\chi^2$ distribution in the mass-inclination plane is shown in Fig.S5. The best fitting dynamical mass (within 1.1" from the centre) is 1.13±0.1 $10^{11}$ M$_\odot$ and the best fitting inclination is 48±3°. Hence the peak circular velocity at a radius of two radial scale lengths is 233±16 km/s. Extrapolating the above mass distribution to larger radii gives 1.28 and 1.49 times greater masses at 1.5" and 2", respectively.

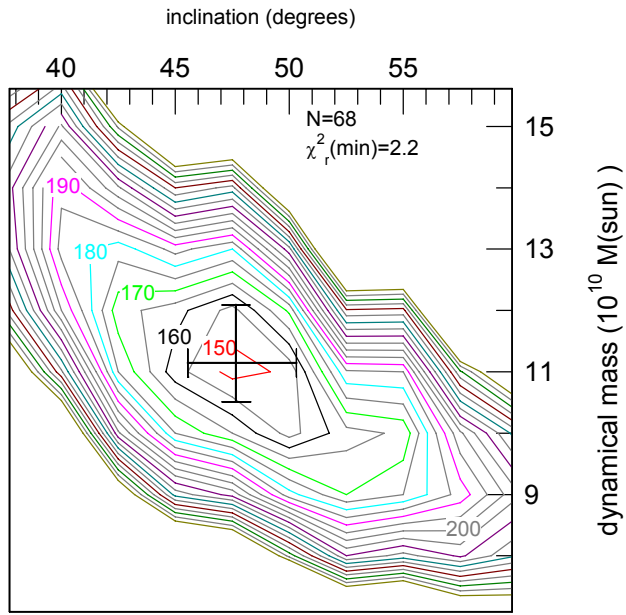

Figure S5. Distribution of $\chi^2$ for the 68 data points shown in Fig.2 (high resolution data only) in the mass-inclination plane. The best fitting dynamical mass (within 1.1", the outer radius covered by data) and inclination and their $1\sigma$ uncertainties is shown as a cross.

Finally, the dispersion distribution along the major axis is most sensitive to local velocity dispersion in excess of the instrumental width. In terms of the z-scale height/z-velocity dispersion model discussed above we find the best fitting FWHM z-thickness of 0.16"±0.03", resulting in $v_c/\sigma(z)$=3±1 at the peak of the rotation curve ($r=2R_{1/e}$). While we found that the 4 fit parameters were obtained in the most straightforward and de-coupled way from the four cuts shown in Figure 2 of the paper, consistent results were obtained from $\chi^2$-minimization of the residual mean velocity and dispersion maps shown in Figure 3 of the paper, after masking out the central parts containing strong residuals of radial motions.